# Predicting A Novel Phase of 2D SiTe$_2$

*Romakanta Bhattarai and Xiao Shen\**

Department of Physics and Materials Science, University of Memphis, Memphis, TN, 38152

\* email: xshen1@memphis.edu

**ABSTRACT**

Layered IV-VI$_2$ compounds often exist in the CdI$_2$ structure. Using the evolution algorithm and first-principles calculations, we predict a novel layered structure of silicon ditelluride (SiTe$_2$) that is more stable than the CdI$_2$ phase. The structure has a triclinic unit cell in its bulk form and exhibits the competition between the Si atoms' tendency to form tetrahedral bonds and the Te atoms' tendency to form hexagonal close-packing. The electronic and vibrational properties of the predicted phase are investigated. The effective mass of electron is small among 2D semiconductors, which is beneficial for applications such as field-effect transistors. The vibrational Raman and IR spectra are calculated to facilitate future experimental investigations.

## I. INTRODUCTION

The investigation and characterization of two dimensional (2D) materials have been increased significantly since the last decade. Synthesis of many novel 2D materials including graphene [1,2], hexagonal boron nitrides[3,4], transition metal dichalcogenides[5,6], phosphorene[7,8], and silicene[9,10] drives rapid progress in the field. These materials attract significant interest because of their intriguing properties different than in their bulk forms due to the reduced dimension, such as quantum confinement, mechanical flexibility, lack of dielectric screening, and large surface areas. These properties are beneficial for a wide range of potential applications in optoelectronics[11,12], chemical sensors[13], photovoltaics[14], energy storage[15], nanoelectronics[16], and many more.

The IV$_x$-VI$_y$ compound family contains a number of materials with layered crystal structures from which 2D crystalline mono- and multi-layers can be obtained. The IV-VI group includes 2D

SnSe[17], SnS, GeS, and GeSe[18], which adopt a puckered layer structure with $C_{2v}$ symmetry. The 2D IV-VI$_2$ materials include SnS$_2$[19] and SnSe$_2$[20], whose bulk forms adopt the CdI$_2$-type crystal structures with P3m1 space group. The Si$_x$-Te$_y$ system is a particularly interesting member of the IV$_x$-VI$_y$ family, as Si is a small group 4 element, and Te is a large group 6 element. The silicon telluride (Si$_2$Te$_3$)[21,22] is the most studied Si$_x$-Te$_y$ compound[23–29]. It is the only known IV$_2$-VI$_3$ material with a layered structure[21,30] and features a unique structural variability because of the orientation of silicon dimers[24]. The other Si$_x$-Te$_y$ compound is silicon ditelluride (SiTe$_2$)[31–34], whose electrical, thermal, and magnetic properties have recently drawn a number of investigations[35–37]. The crystal structure of SiTe$_2$ is identified be the CdI$_2$-type[34,38] as other 2D IV-VI$_2$ materials, although it was suggested that it may also exist in the Si$_2$Te$_3$ structure with Si deficiency[39].

In this paper, we use results from the evolutionary algorithm and first-principles calculations to predict a new layered crystal structure of SiTe$_2$ that is more stable than the CdI$_2$-type structure. The predicted bulk structure has a triclinic unit cell. The atomic structure indicates the competition between the Si atoms' tendency to form tetrahedral bonds and the Te atoms' tendency to form hexagonal close-packing. The material has a low electron effective mass and anisotropic hole effective mass that can be beneficial for potential applications. The Raman and IR spectra are calculated, which can be useful for future experimental investigations.

## II. COMPUTATIONAL METHODS

**Evolutionary Algorithm.** The global search of stable crystal structures of bulk SiTe$_2$ was done using the evolutionary algorithm as implemented in the universal crystal structure predictor USPEX[40,41]. The initial population (first generation) consists of 30 structures generated randomly by using the space group symmetry under fixed chemical composition Si:Te=1:2. The number of structures in each generation is kept constant. The genetic evolutionary could stop if the best structure did not change for ten generations. We use 60% of the current generation to produce the next generation. In total, 50% of the generation was produced by heredity, 30% of generation produced randomly from space groups, and 20% of the generation produced by soft mutations in each generation afterward. The external pressure is set to zero. The structural relaxations and total energy calculations are carried using density functional theory (DFT) as implemented in the VASP

(Vienna Ab initio Simulation Package) code[42]. The pseudopotentials are constructed under the projected augmented wave (PAW) method[43]. We used the Perdew-Burke-Ernzerhof (PBE) form of exchange-correlation functional[44] under generalized gradient approximation (GGA). The global break conditions for the electronic and ionic steps are $10^{-5}$ eV and $10^{-4}$ eV/Å, respectively. During the evolutionary search of structures, the maximum kinetic energy cutoff for the plane-wave basis is 320 eV.

**Density Functional Theory Calculations.** After we found the most stable structure of bulk $SiTe_2$, we construct the monolayer and investigate the electronic properties of both bulk and the monolayer through DFT and hybrid DFT calculations. These calculations are carried out using the VASP code and the same pseudopotentials as in the evolutionary algorithm search. We use the PBE functional for DFT calculations and hybrid HSE06 functional[45,46] for hybrid DFT calculations. The convergence conditions for the electronic and ionic steps are $10^{-9}$ eV and $10^{-8}$ eV, respectively. The kinetic energy cutoff for the plane-wave basis set was 500 eV. Brillouin zones for bulk and monolayer $SiTe_2$ respectively were sampled at 13×13×7 and 13×13×1 k-point grids centered at the Γ point. The spin-orbit coupling effects are considered for the band structure calculations. For the model of the monolayer structure, a vacuum of 13.7 Å is inserted between the periodic replicas of the monolayer to avoid artificial interactions. To confirm the dynamical stability, we performed the phonon calculations of a 5×5×1 supercell of the monolayer using the finite displacement method as implemented in the Phonopy program[47]. To further verify the dynamical stability, we carried out ab-initio molecular dynamic simulations (AIMD) in a 4×4×1 supercell of the monolayer at 500 K. The time step was set to 2 fs with a total simulation time of 20 ps.

**Raman and IR spectra**. We carried out density functional perturbation theory (DFPT)[48] calculations as implemented in Quantum Espresso package[49] to obtain the Raman and IR spectra of bulk and monolayer of $SiTe_2$. The calculations use the Perdew-Zunger functional[50] under the local density approximation and the norm-conserving pseudopotential[51]. Reciprocal space was sampled with a k-point grid of 13×13×7 for bulk and 13×13×1 for monolayer centered at the Γ point. The cutoff energy of the plane-wave basic is 80 Ry. The Raman and IR spectra calculations start from the charge densities that are converged to $10^{-14}$ Ry in total energy.

## III. RESULTS AND DISCUSSION

**Structures.** A total of 382 stable crystalline structures are generated by the evolution algorithm. In Figure 1a we plotted the energy per atom of each of the generated structures. We obtain the lowest energy structure with the total energy of -3.88 eV/atom. The top and side views of the corresponding structure are shown in Figures 2a and 2b. The structure features a triclinic crystal lattice with a space group P1. The primitive unit cell consists of three atoms (one Si atom and two Te atoms). The lattice parameters of this lowest energy of $SiTe_2$ are shown in Table I. The evolutionary algorithm search also found the $CdI_2$-type structure of $SiTe_2$ that was previously reported[34,38] with the total energy of -3.81 eV/atom. Thus the new predicted structure is energetically more stable than the common $CdI_2$-type structure that has been reported in $SiTe_2$.

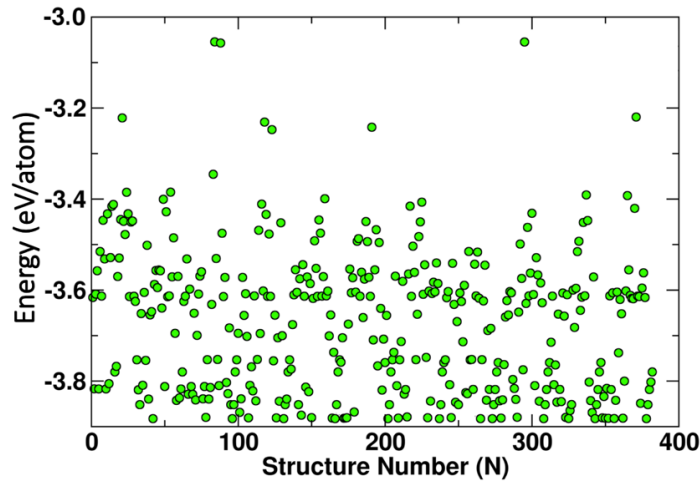

Figure 1: Energies for all the generated structures of $SiTe_2$.

It is interesting to discuss why this structure exists in $SiTe_2$ and is more stable than the CdI2-type structure. In Figure 2c, we show the local bonding structure of Si in this new structure. The Si atom is clearly tetrahedrally bonded, which indicates covalent bonding from sp3 hybridization of Si 3s and 3p orbitals. Meanwhile, in the $CdI_2$-type structure, the cations are six-coordinated (Figure 2d). Given the strong tendency of Si to form such covalent bonds, it is reasonable to state that forming six-coordinated Si is associated with an energy penalty compared with four-coordinated Si. On the other hand, the tetrahedral around the Si atoms are strongly distorted. As shown in Table II, there is a large variety of the Te-Te distances in the tetrahedral surrounding the Si atom, ranging from 3.945 Å to 4.516 Å. The reason for this distortion is that due to its large size, the Te atom prefers hexagonal close-packed (HCP) structure, which is not fully compatible with the tetrahedral bonding of Si. The competing requirements lead to the distorted bonding around Si atoms along with a slightly distorted HCP packing of Te, as can be seen in Figures 2e and 2f.

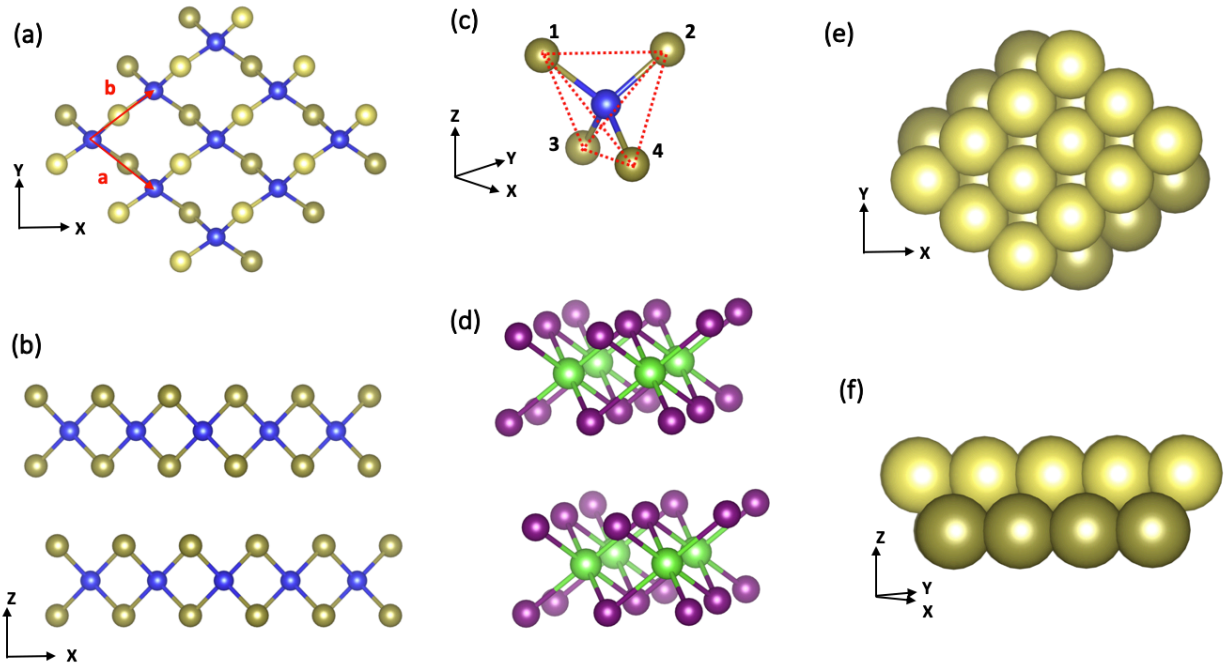

Figure 2: (a) Top and (b) side views of the crystal structure of SiTe$_2$ with unit lattice vector marked by solid red lines. Te and Si atoms are represented by tan and blue colors, respectively. The Te atoms in the upper layer are shown in the brighter tan color. (c) The bonding configuration near a Si atom. (d) The CdI$_2$ structure. (e) Top and (f) side views of the space-filling plot showing the packing of Te atoms.

Table I: Structural parameters of predicted bulk and monolayer SiTe$_2$.

|  | a (Å) | b (Å) | c (Å) | α (°) | β (°) | γ (°) | $d_{Si-Te}$ (Å) |
|---|---|---|---|---|---|---|---|
| Bulk | 3.945 | 3.945 | 7.076 | 83.68 | 83.77 | 75.98 | 2.561 |
| Monolayer | 3.941 | 3.941 | N/A | N/A | N/A | 75.88 | 2.558 |

Table II: Nearest neighbor Te-Te distances in bulk SiTe$_2$.

| $d_{Te1-Te2}$ (Å) | $d_{Te1-Te3}$ (Å) | $d_{Te1-Te4}$ (Å) | $d_{Te2-Te3}$ (Å) | $d_{Te2-Te4}$ (Å) | $d_{Te3-Te4}$ (Å) |
|---|---|---|---|---|---|
| 3.945 | 4.070 | 4.516 | 4.502 | 4.070 | 3.945 |

**Dynamical Stability.** In order to determine whether the predicted structure is dynamical stable, we calculated the phonon dispersion of the monolayer SiTe$_2$ along the high symmetry lines of Brillouin zones using the finite displacement method. The phonon spectrum (Figure 3a) shows no significant negative frequencies, therefore indicating the structure is dynamically stable. The very small negative frequencies in the acoustic mode near the Γ point are numerical artifacts due to the use of a finite FFT grid. The dynamical stability is further examined by the ab-initio molecular dynamic (AIMD) simulation at 500 K. As shown in Figure 3b, the total energy fluctuates during the simulation without any sudden drop of the energy that would indicate a phase transition. The snapshot confirms that the structure remains stable during the AIMD simulations.

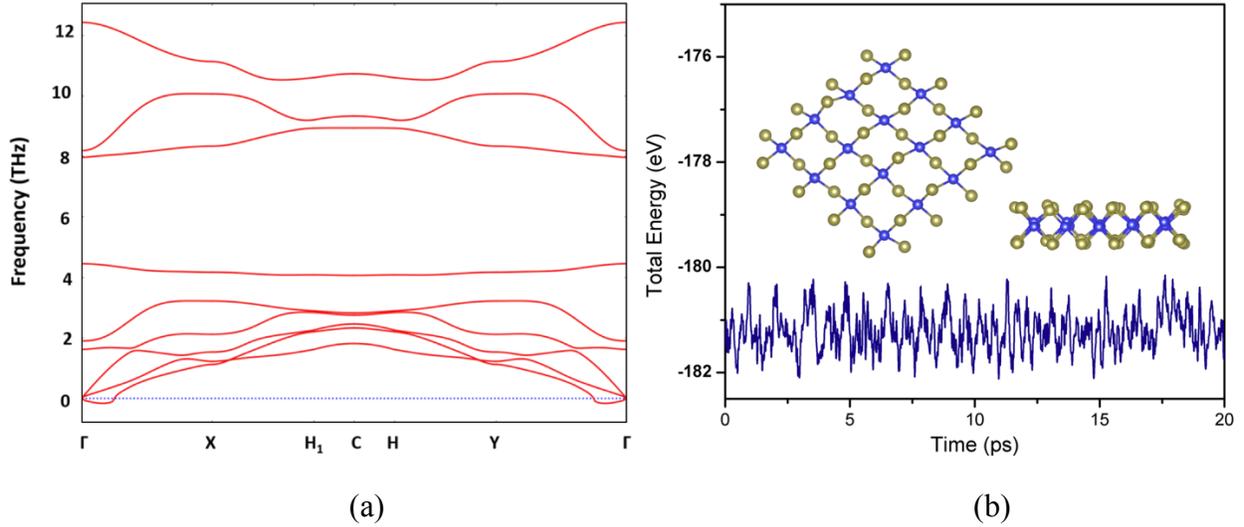

(a)  (b)

Figure 3: (a) Phonon band structures of the monolayer of SiTe$_2$. (b)A snapshot of for AIMD simulation at 500 K and the variation of total energy in AIMD simulation during a timescale of 20 ps.

**Electronic Structure.** To understand the electronic properties of SiTe$_2$, we studied the band structures of both the bulk and monolayer along the high symmetry lines of Brillouin zones using the DFT method. Figures 4(a) and 4(b) represent the electronic band structures of bulk and monolayer SiTe$_2$, taking into account the spin-orbit interaction. The bandgap is 0.206 eV for bulk and 0.552 eV for the monolayer. In bulk SiTe$_2$, the valence band maximum (VBM) lies between the $\Gamma$ and Z and the conduction band minimum (CBM) is at $\Gamma$. Meanwhile, both VBM and CBM are located at the $\Gamma$ point in the monolayer. These results indicate that the bulk and monolayer SiTe$_2$ are indirect and direct bandgap semiconductors, respectively. The bandgap in the monolayer is higher than in bulk because of the quantum confinement effect. Since DFT is known to underestimate the bandgap, we used the hybrid DFT method with the HSE06 functional. Figure 5 below represents the band structures of SiTe$_2$ under the HSE06 approach. The band structures are similar to the DFT band, except that the band gaps for bulk and monolayer have increased to 0.831 and 1.222 eV, respectively.

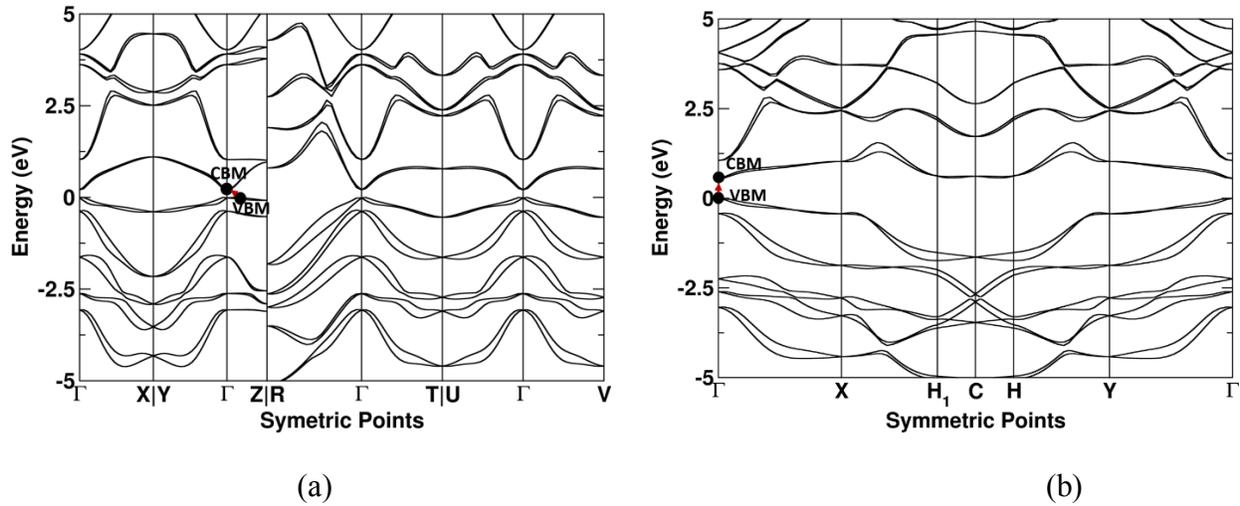

(a)　　　　　　　　　　　　　　　　　　　(b)

Figure 4: Electronic band structures of (a) bulk and (b) monolayer SiTe$_2$ under DFT approach showing the indirect and direct band gaps taking the spin-orbit interaction into account.

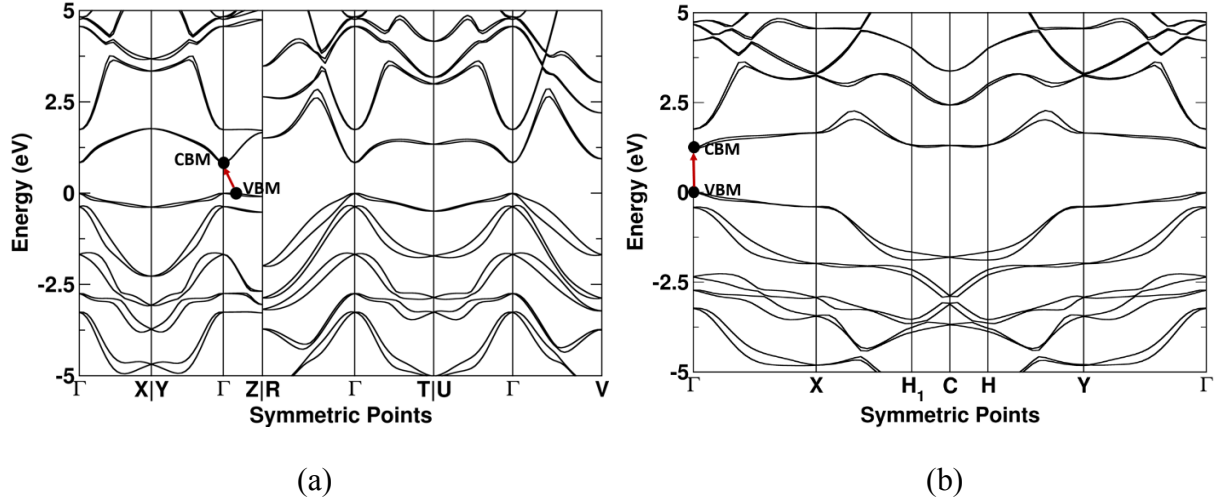

(a)                                       (b)

Figure 5: Electronic band structures of (a) bulk and (b) monolayer SiTe$_2$ with HSE06 functional.

The effective masses of the electrons and holes in SiTe$_2$ are obtained by fitting the DFT bands near the CBM and VBM using the relationship $m^* = \hbar^2/(\frac{\partial^2 E}{\partial k^2})$. The results are shown in Table III. The electron effective mass bulk is 0.116 $m_0$ in the x-direction, which is comparable to black phosphorous ($m^* = 0.12\ m_0$)[52] and four times smaller than MoS$_2$ ($m^* = 0.45\ m_e$)[53]. The low electron effective mass suggests high electron mobility, which is beneficial for applications such as field-effect transistors. An in-plane anisotropy is observed for the holes, where the effective mass along the y-direction is 3 to 4 times the value along the x-direction. Such anisotropy may be beneficial for designing devices such as angle-dependent optoelectronic applications[54].

Table III: The carrier effective masses in SiTe$_2$.

|  | carrier | xx ($m_0$) | yy ($m_0$) | zz ($m_0$) |
|---|---|---|---|---|
| Bulk | e | 0.116 | 0.163 | 0.229 |
|  | h | 0.160 | 0.667 | 4.329 |
| Monolayer | e | 0.177 | 0.227 | N/A |
|  | h | 0.262 | 0.623 | N/A |

Figure 6a shows the partial and total densities of states (DOS) of bulk SiTe$_2$ from HSE calculations. The conduction bands consist both the p orbitals of Si atoms and the p orbitals of Te atoms. Meanwhile, the valence bands are mainly due to the p orbitals of Te atoms. To better understand the contribution of each atomic orbitals to the band structures, we analyze the wave functions at band edges. Figure 6b and 6c represent the square of the wave functions at CBM and VBM. It can clearly be seen the s orbitals of Si atoms along with $p_x+p_y$ orbitals of Te atoms play a major role at CBM, while the VBM mainly consists of $p_x+p_y$ orbitals of Te atom. It is interesting to note that the $p_x+p_y$ orbitals from adjacent Te atoms tend to align with the Te-Te direction, which indicates the σ type bonding between Te atoms.

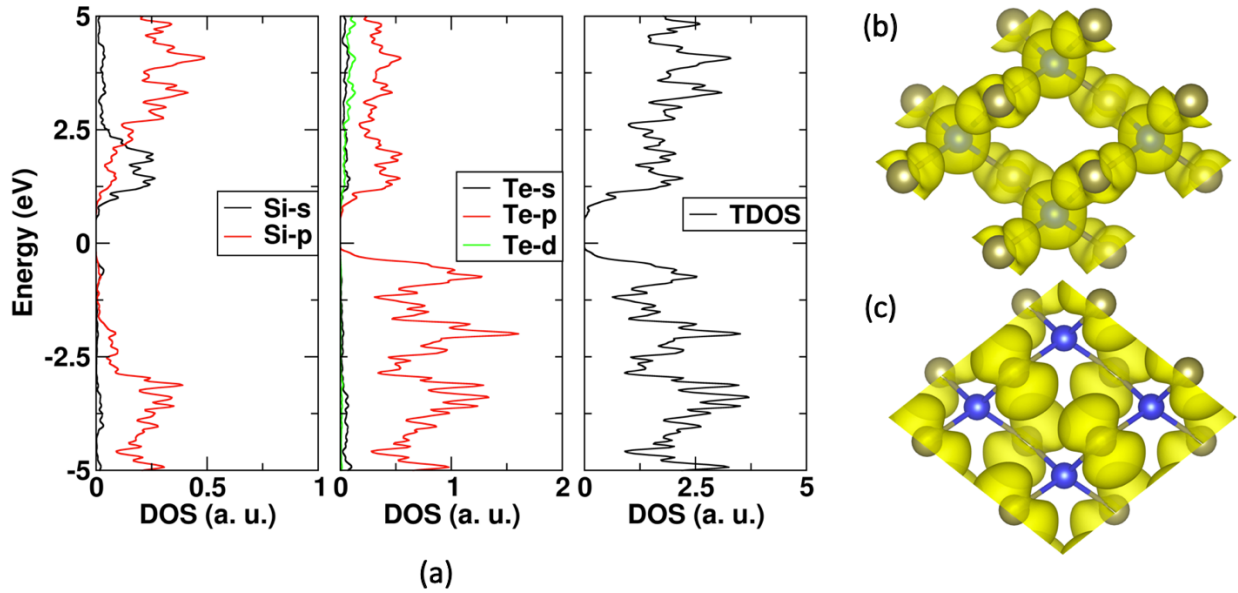

Figure 6: (a) Partial and total DOS of SiTe$_2$ from HSE calculations. Charge density plots at CBM (b) and VBM (c).

**Raman and IR spectra.** Vibrational spectroscopy is widely used to characterize 2D materials. To facilitate future experimental investigation of the predict phase of SiTe$_2$, we present the calculated Raman spectra of bulk and monolayer SiTe$_2$ with the respective vibrational modes in Figure 7. We found the bulk SiTe$_2$ has three major Raman peaks at 144.06 cm$^{-1}$, 261.73 cm$^{-1}$, and 409.90 cm$^{-1}$, whereas the monolayer SiTe$_2$ has two major Raman peaks at 147.35 cm$^{-1}$ and 413.88 cm$^{-1}$. The 144.06/147.35 cm$^{-1}$ peak corresponds to the out-of-plane vibration mode from the Te atoms, while the 409.90/413.88 cm$^{-1}$ peak corresponds to the out-of-plane vibration mode from the Si atoms. The peak at 261.73 cm$^{-1}$ in bulk SiTe$_2$ corresponds to the in-plane vibration of Si atoms.

This mode has negligible Raman intensity in the monolayers while being significant in bulk, indicating the effect of interlayer coupling.

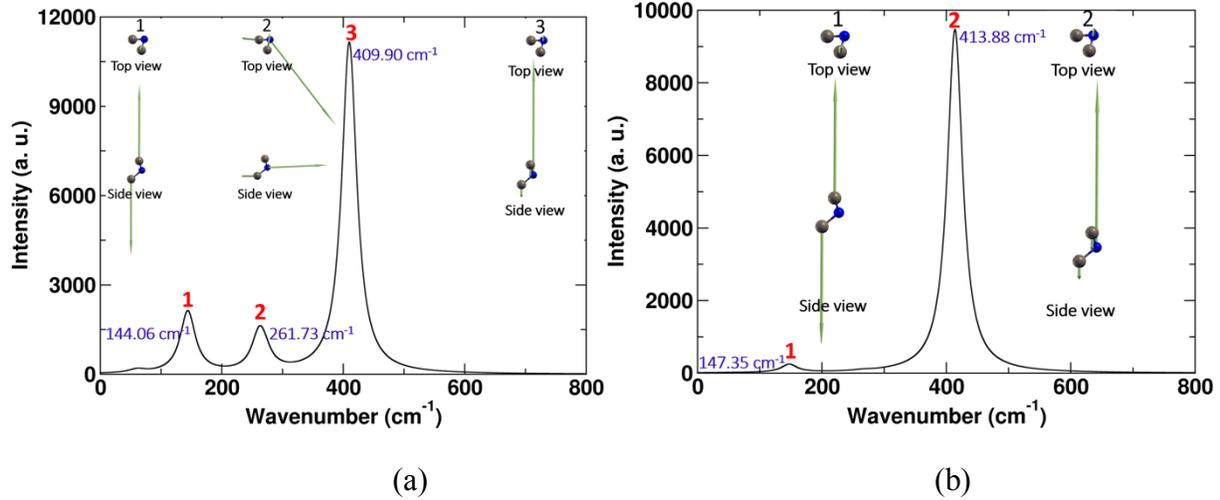

Figure 7: Raman spectra of (a) bulk and (b) monolayer SiTe$_2$ showing major peaks.

Figure 8 shows the calculated IR spectra. The bulk SiTe$_2$ has two major peaks at 269.89 cm$^{-1}$ and 409.00 cm$^{-1}$. The former corresponds to an in-plane vibration mode of Si atoms that is orthogonal to the Raman-active in-plane Si vibration mode at 261.73 cm$^{-1}$, while the latter is the same as the Raman-active out-of-plane Si vibration mode. The monolayer SiTe$_2$ has only one major Raman peaks at 271.79 cm$^{-1}$. The mode around 410 cm$^{-1}$ that is IR active in bulk is inactive in the monolayers, again indicating the effect of interlayer coupling. This feature may be useful for identifying the thickness of the SiTe$_2$ layers in experiments.

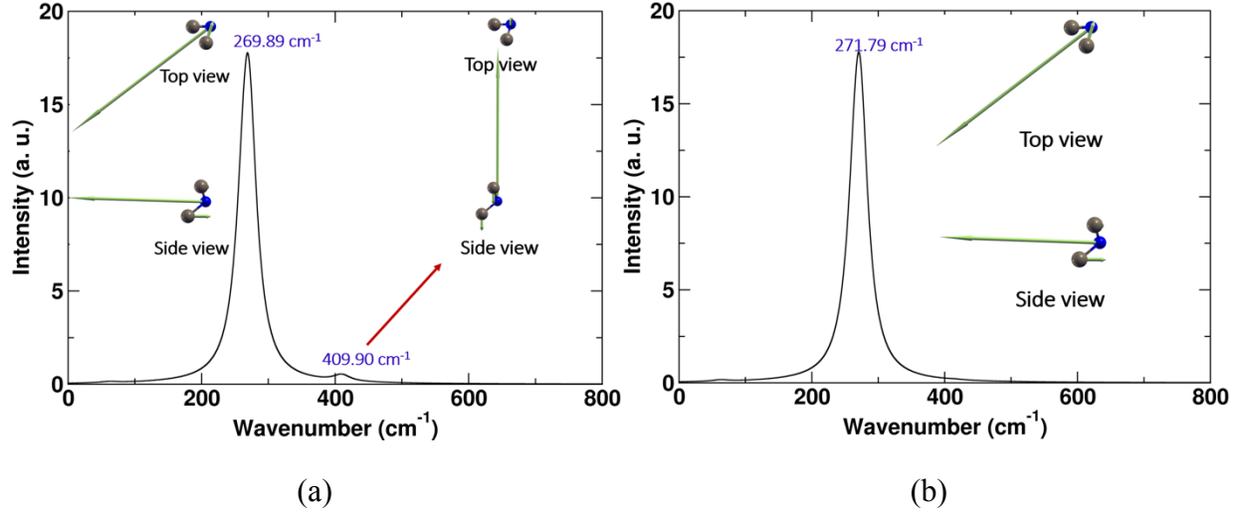

Figure 8: IR spectra of (a) bulk and (b) monolayer SiTe$_2$ showing major peaks.

## IV. SUMMARY

In summary, we use the evolution algorithm coupled with the first-principles calculations to predict a novel 2D layered structure of silicon telluride (SiTe$_2$), which is different from the known layered CdI2-type structure in IV-VI$_2$ materials. The structure features a distorted tetrahedral bonding for Si atoms and a distorted hexagonal close pack stacking of Te atoms, indicating that structure forms as a compromise of these two competing requirements. We confirm the dynamical stability of the structure using both phonon dispersion and the AIMD simulations. Electronic properties are investigated by both the DFT and hybrid DFT methods. The new SiTe$_2$ phase is a semiconductor with an indirect bandgap in its bulk form and a direct bandgap in its monolayer form. It features electron effective mass of 0.12 m$_0$, which is low among 2D semiconductors. The Raman and IR spectra for the bulk and monolayers are also predicted.

## ACKNOWLEDGEMENT

This work was supported by the National Science Foundation under Grant No. DMR 1709528 and by the Ralph E. Powe Jr. Faculty Enhancement Awards from Oak Ridge Associated Universities (ORAU). Computational resources were provided by the NSF XSEDE grant number

**REFERENCES**

1. Berger, C. *et al.* Ultrathin Epitaxial Graphite: 2D Electron Gas Properties and a Route toward Graphene-based Nanoelectronics. *J. Phys. Chem. B* **108**, 19912–19916 (2004).

2. Novoselov, K. S. *et al.* Electric Field Effect in Atomically Thin Carbon Films. *Science* **306**, 666–669 (2004).

3. Pacilé, D., Meyer, J. C., Girit, Ç. Ö. & Zettl, A. The two-dimensional phase of boron nitride: Few-atomic-layer sheets and suspended membranes. *Appl. Phys. Lett.* **92**, 133107 (2008).

4. Han, W.-Q., Wu, L., Zhu, Y., Watanabe, K. & Taniguchi, T. Structure of chemically derived mono- and few-atomic-layer boron nitride sheets. *Appl. Phys. Lett.* **93**, 223103 (2008).

5. Gordon, R. A., Yang, D., Crozier, E. D., Jiang, D. T. & Frindt, R. F. Structures of exfoliated single layers of WS2, MoS2, and MoSe2 in aqueous suspension. *Phys. Rev. B* **65**, 125407 (2002).

6. Mak, K. F., Lee, C., Hone, J., Shan, J. & Heinz, T. F. Atomically Thin MoS2 : A New Direct-Gap Semiconductor. *Phys. Rev. Lett.* **105**, 136805 (2010).

7. Li, L. *et al.* Black phosphorus field-effect transistors. *Nat. Nanotechnol.* **9**, 372–377 (2014).

8. Liu, H. *et al.* Phosphorene: An Unexplored 2D Semiconductor with a High Hole Mobility. *ACS Nano* **8**, 4033–4041 (2014).

9. Cahangirov, S., Topsakal, M., Aktürk, E., Şahin, H. & Ciraci, S. Two- and One-Dimensional Honeycomb Structures of Silicon and Germanium. *Phys. Rev. Lett.* **102**, 236804 (2009).

10. Vogt, P. *et al.* Silicene: Compelling Experimental Evidence for Graphenelike Two-Dimensional Silicon. *Phys. Rev. Lett.* **108**, 155501 (2012).

11. Wang, Q. H., Kalantar-Zadeh, K., Kis, A., Coleman, J. N. & Strano, M. S. Electronics and optoelectronics of two-dimensional transition metal dichalcogenides. *Nat. Nanotechnol.* **7**, 699–712 (2012).

12. Wilson, J. A. & Yoffe, A. D. The transition metal dichalcogenides discussion and interpretation of the observed optical, electrical and structural properties. *Adv. Phys.* **18**, 193–335 (1969).

13. Anichini, C. *et al.* Chemical sensing with 2D materials. *Chem. Soc. Rev.* **47**, 4860–4908 (2018).

14. Velusamy, D. B. *et al.* Flexible transition metal dichalcogenide nanosheets for band-selective photodetection. *Nat. Commun.* **6**, 1–11 (2015).

15. Anasori, B., Lukatskaya, M. R. & Gogotsi, Y. 2D metal carbides and nitrides (MXenes) for energy storage. *Nat. Rev. Mater.* **2**, 16098 (2017).

16. Shafique, A., Samad, A. & Shin, Y.-H. Ultra low lattice thermal conductivity and high carrier mobility of monolayer SnS2 and SnSe2: a first principles study. *Phys. Chem. Chem. Phys.* **19**, 20677–20683 (2017).

17. Li, L. *et al.* Single-Layer Single-Crystalline SnSe Nanosheets. *J. Am. Chem. Soc.* **135**, 1213–1216 (2013).

18. Fei, R., Li, W., Li, J. & Yang, L. Giant piezoelectricity of monolayer group IV monochalcogenides: SnSe, SnS, GeSe, and GeS. *Appl. Phys. Lett.* **107**, 173104 (2015).

19. Sun, Y. *et al.* Freestanding Tin Disulfide Single-Layers Realizing Efficient Visible-Light Water Splitting. *Angew. Chem. Int. Ed.* **51**, 8727–8731 (2012).

20. Zhou, X. *et al.* Photodetectors: Ultrathin SnSe2 Flakes Grown by Chemical Vapor Deposition for High-Performance Photodetectors (Adv. Mater. 48/2015). *Adv. Mater.* **27**, 8119–8119 (2015).

21. Ploog, K., Stetter, W., Nowitzki, A. & Schonherr, E. Crystal Growth and Structure Determination of Silicon Telluride Si2Te3. *Mater. Res. Bull.* **11**, 1147–1154 (1976).

22. Zwick, U. & Rieder, K. H. Infrared and Raman Study of Si2Te3. *Z Phys. B* **25**, 319 (1976).

23. Keuleyan, S., Wang, M., Chung, F. R., Commons, J. & Koski, K. J. A Silicon-Based Two-Dimensional Chalcogenide: Growth of Si2Te3 Nanoribbons and Nanoplates. *Nano Lett.* **15**, 2285–2290 (2015).

24. Shen, X., Puzyrev, Y. S., Combs, C. & Pantelides, S. T. Variability of structural and electronic properties of bulk and monolayer Si2Te3. *Appl. Phys. Lett.* **109**, 113104 (2016).

25. Wu, K., Chen, J., Shen, X. & Cui, J. Resistive switching in Si2Te3 nanowires. *AIP Adv.* **8**, 125008 (2018).

26. Kwak, J., Thiyagarajan, K., Giri, A. & Jeong, U. Au-Assisted catalytic growth of Si2Te3 plates. *J. Mater. Chem. C* **7**, 10561–10566 (2019).

27. Chen, J., Wu, K., Shen, X., Hoang, T. B. & Cui, J. Probing the dynamics of photoexcited carriers in Si2Te3 nanowires. *J. Appl. Phys.* **125**, 024306 (2019).

28. Bhattarai, R. & Shen, X. Ultra-high mechanical flexibility of 2D silicon telluride. *Appl. Phys. Lett.* **116**, 023101 (2020).

29. Bhattarai, R., Chen, J., Hoang, T. B., Cui, J. & Shen, X. Anisotropic Optical Properties of 2D Silicon Telluride. *MRS Adv.* 1–9 (2020) doi:10.1557/adv.2020.186.

30. Gregoriades, P. E., Bleris, G. L. & Stoemenos, J. Electron diffraction study of the Si2Te3 structural transformation. *Acta Crystallogr. B* **39**, 421–426 (1983).

31. Weiss, A. & Weiss, A. Siliciumchalkogenide. IV. Zur Kenntnis von Siliciumditellurid. *Z. Für Anorg. Allg. Chem.* **273**, 124–128 (1953).

32. Rau, J. W. & Kannewurf, C. R. Intrinsic absorption and photoconductivity in single crystal SiTe2. *J. Phys. Chem. Solids* **27**, 1097–1101 (1966).

33. Lambros, A. P. & Economou, N. A. The Optical Properties of Silicon Ditelluride. *Phys. Status Solidi B* **57**, 793–799 (1973).

34. Taketoshi, K. & Andoh, F. Structural Studies on Silicon Ditelluride (SiTe2). *Jpn. J. Appl. Phys.* **34**, 3192 (1995).

35. Prajapat, C. L. *et al.* Transport and magnetic properties of SiTe2. *AIP Conf. Proc.* **1832**, 130011 (2017).

36. Kandemir, A., Iyikanat, F. & Sahin, H. Stability, electronic and phononic properties of β and 1T structures of SiTex(x= 1, 2) and their vertical heterostructures. *J. Phys. Condens. Matter* **29**, 395504 (2017).

37. Wang, Y., Gao, Z. & Zhou, J. Ultralow lattice thermal conductivity and electronic properties of monolayer 1T phase semimetal SiTe2 and SnTe2. *Phys. E Low-Dimens. Syst. Nanostructures* **108**, 53–59 (2019).

38. Mishra, R. *et al.* Evidences of the existence of SiTe2 crystalline phase and a proposed new Si–Te phase diagram. *J. Solid State Chem.* **237**, 234–241 (2016).

39. Göbgen, K. C., Steinberg, S. & Dronskowski, R. Revisiting the Si–Te System: SiTe2 Finally Found by Means of Experimental and Quantum-Chemical Techniques. *Inorg. Chem.* **56**, 11398–11405 (2017).

40. Lyakhov, A. O., Oganov, A. R., Stokes, H. T. & Zhu, Q. New developments in evolutionary structure prediction algorithm USPEX. *Comput. Phys. Commun.* **184**, 1172–1182 (2013).

41. Oganov, A. R., Lyakhov, A. O. & Valle, M. How Evolutionary Crystal Structure Prediction Works—and Why. *Acc. Chem. Res.* **44**, 227–237 (2011).

42. Kresse, G. & Furthmüller, J. Efficient iterative schemes for ab initio total-energy calculations using a plane-wave basis set. *Phys. Rev. B* **54**, 11169–11186 (1996).

43. Blöchl, P. E. Projector augmented-wave method. *Phys. Rev. B* **50**, 17953–17979 (1994).

44. Perdew, J. P., Burke, K. & Ernzerhof, M. Generalized Gradient Approximation Made Simple. *Phys. Rev. Lett.* **77**, 3865–3868 (1996).

45. Heyd, J., Scuseria, G. E. & Ernzerhof, M. Hybrid functionals based on a screened Coulomb potential. *J. Chem. Phys.* **118**, 8207–8215 (2003).

46. Heyd, J., Scuseria, G. E. & Ernzerhof, M. Erratum: "Hybrid functionals based on a screened Coulomb potential" [J. Chem. Phys. 118, 8207 (2003)]. *J. Chem. Phys.* **124**, 219906 (2006).

47. Togo, A. & Tanaka, I. First principles phonon calculations in materials science. *Scr. Mater.* **108**, 1–5 (2015).

48. Gonze, X. & Vigneron, J.-P. Density-functional approach to nonlinear-response coefficients of solids. *Phys. Rev. B* **39**, 13120–13128 (1989).

49. Giannozzi, P. *et al.* QUANTUM ESPRESSO: a modular and open-source software project for quantum simulations of materials. *J. Phys. Condens. Matter* **21**, 395502 (2009).

50. Perdew, J. P. & Zunger, A. Self-interaction correction to density-functional approximations for many-electron systems. *Phys Rev B* **23**, 5048–5079 (1981).

51. QUANTUMESPRESSO - QUANTUMESPRESSO. http://www.quantum-espresso.org/.

52. Qiao, J., Kong, X., Hu, Z.-X., Yang, F. & Ji, W. High-mobility transport anisotropy and linear dichroism in few-layer black phosphorus. *Nat. Commun.* **5**, 4475 (2014).

53. Peelaers, H. & Van de Walle, C. G. Effects of strain on band structure and effective masses in MoS2. *Phys Rev B* **86**, 241401 (2012).

54. Li, L. *et al.* Emerging in-plane anisotropic two-dimensional materials. *InfoMat* **1**, 54–73 (2019).